\documentclass[aps,prb,twocolumn,groupedaddress,floatfix]{revtex4}
\usepackage{graphics,graphicx}
\usepackage{color}
\usepackage{amssymb,amsmath}
\usepackage{bm}

\newcommand{\req}[1]{Eq.~(\ref{#1})}
\newcommand{\tdm}{{\tilde \Delta}_{m}}
\newcommand{\tem}{{\tilde \omega}_{m}}
\newcommand{\barD}{{\bar \Delta}_{m}}
\newcommand{\vare}{\omega}
\newcommand{\Del}{{\mbox{\footnotesize $\Delta$}}}
\newcommand{\ag}{\zeta}
\newcommand{\vpi}{{\bf \pi}}
\newcommand{\be}{\begin{equation}}
\newcommand{\ee}{\end{equation}}
\newcommand{\bea}{\begin{eqnarray}}
\newcommand{\eea}{\end{eqnarray}}
\begin{document}
\unitlength = 1mm
%~~~~~~~~~~~~~~~~~~~~~~~~~~~~~~~~~~~~~~~~~~~~~~~~~
\title{Superfluid density and penetration depth in Fe-pnictides}
\author{A.~B.~Vorontsov, M.~G.~Vavilov, and A.~V.~Chubukov}
\affiliation{Department of Physics,
             University of Wisconsin, Madison, Wisconsin 53706, USA}

\date{March 15, 2009}

\begin{abstract}
We consider the superfluid density $\rho_s (T)$ in a two-band superconductor with sign-changing
extended $s$-wave symmetry ($s^+$) in the presence of non-magnetic impurities and apply
the results to Fe-pnictides. We show that the behavior of the superfluid
density is essentially the same as in an ordinary $s$-wave superconductor with
magnetic impurities. We show that, for moderate to strong inter-band impurity scattering,
$\rho_s (T)$ behaves as a power-law $T^n$ with $n\approx 1.6\div2$
over a wide range of $T$. We argue that the power-law behavior is consistent
with  recent experiments on the penetration depth $\lambda (T)$
in doped BaFe$_2$As$_2$, but disagree quantitatively
with the data on LaFePO.
\end{abstract}
%~~~~~~~~~~~~~~~~~~~~~~~~~~~~~~~~~~~~~~~~~~~~~~~~~~~~~~~~~~~~~~~~~~~~~~~~~~~~~
\maketitle
%~~~~~~~~~~~~~~~~~~~~~~~~~~~~~~~~~~~~~~~~~~~~~~~~~~~~~~~~~~~~~~~~~~~~~~~~~~~~~

{\it Introduction.}~~~
Recent discovery of iron-based pnictide superconductors
instigated massive theoretical and experimental research effort
aimed at unveiling fundamental properties of these materials.
Both oxygen containing `1111's materials
(La,Nd,Pr,Sm)FeAsO, and oxygen free `122's
(Ca,Ba,Sr)Fe$_2$As$_2$ have high potential for the applications and may create
a breakthrough in the field of superconductivity (SC).\cite{NormanPhysNews}

One of the central and still unsettled issues is the symmetry of the SC gap.
An ordinary $s$-wave superconductivity due to phonons has been deemed unlikely
because of too small electron-phonon coupling,\cite{phonons}
suggesting that the SC pairing is of electronic origin.
Electronic structure of pnictides shows pairs of
small hole and electron pockets centered at $(0,0)$ and
$(\pi,\pi)$, respectively in the \emph{folded} Brillouin
zone.\cite{phonons,lda,arpes} Most of parent compounds display an SDW
order with momentum at, or near $(\pi,\pi)$, and an early
scenario was the pairing mediated by antiferromagnetic spin
fluctuations.\cite{mazin} For pnictide geometry, this mechanism
yields an extended $s$-wave gap which
 changes sign between hole
and electron pockets but remains approximately uniform
along  either of them (an $s^+$ gap).
 The $s^+$ gap symmetry has been found in weak
coupling studies of two-band\cite{chubukov08} and
five-band~\cite{d_h_lee} ``$g$-ology''
models of interacting low-energy fermions.
Some other studies, however, found a
gap with extended $s$-wave symmetry in the {\it unfolded} Brillouin
zone.\cite{scal} Such gap has no nodes on the hole Fermi surface (FS), but has
four nodes on  the electron FS, like a $d$-wave gap in the cuprates.  The
uncertainty arises from the fact that in pnictides there is a competition between
the inter-pocket interaction with large momentum transfer and intra-pocket
repulsion. When inter-pocket interaction is stronger, the system likely
develops a  sign-changing $s-$wave gap without nodes; when intra-pocket
repulsion is stronger, the system develops an extended $s-$wave gap with nodes
to minimize the effect of intra-pocket repulsion.\cite{new}

>From experimental perspective, the situation is also unclear. Andreev
spectroscopy\cite{andr} and ARPES measurements\cite{ding} are consistent with
the gap without nodes. NMR and Knight shift measurements\cite{nakai} were
originally interpreted as evidence of a gap with nodes, but it turns out that
the data can be fitted equally well by a dirty $s^+$
superconductor.\cite{chubukov08,parker}  The situation is further complicated
by the fact that the two hole FS are of different sizes, and have different
gaps.\cite{diffgap}

To truly distinguish between $s^+$ gap and a gap with the nodes one should
go to very low temperatures. Recently, two groups
reported measurements of the penetration depth $\lambda (T)$ down to
$10^{-2} \; T_c$.  The data seem to point into different
directions. Ames group reported the data on Co-and K-doped
BaFe$_2$As$_2$ (Ref.~\onlinecite{prozorov08}) and demonstrated that down to the
lowest $T$ and for all dopings the $T$ dependence of
$\lambda (T) = \lambda (0) + \Del\lambda (T)$
can be fitted by $\Del \lambda (T) \propto T^2$.
 In SmFeAsO$_{1-x}$F$_x$\cite{bristol} and PrFeAsO\cite{Hashimoto1111}
penetration depth seems to have an exponential
temperature dependence at low $T$, consistent with the gap without
nodes. At the same time, in another 1111 material, LaFePO,
the Bristol group has found
$\Del \lambda \propto T^{1.2}$ down to the lowest
temperatures.\cite{bristol_1}
FS in this material has been reconstructed from
magneto-oscillation measurements\cite{coldea} and consists of
weakly corrugated small-size cylinders making it unlikely
that either hole or electron FSs extend to the points where
$s^+$ order parameter changes sign.

The penetration depth in the clean limit  was considered in
Ref.~\onlinecite{YNagai08} for an $s^+$  gap and in Ref.~\onlinecite{Parish}
for several other gap symmetries.  For a gap without nodes, $\Del \lambda (T)$
is obviously exponential at low $T$, for a gap with nodes it is linear in $T$.
In this paper we discuss to what extent the existing data for $\lambda (T)$
can be described by a dirty SC with  an $s^+$ gap symmetry, and non-magnetic
impurities.  We argue that the $T^2$ behavior observed by Ames group in 122
materials can be fitted over a wide temperature range for various dopings and
in this respect the results for the penetration depth are not in conflict with
ARPES and other measurements which show SC gap without nodes. The existence of
the two different gaps on the two hole FS makes the agreement between
experiments even better.  On the other hand, the linear in $T$ behavior
observed in LaFePO is not reproduced, and it is more likely that the gap in this
material has nodes, as suggested in Ref.~\onlinecite{scal}.

%The input parameters for our fits are the ratio $\Gamma_{\vpi}/T_{c0}$
%($\Gamma_{\vpi}$ is the inter-band scattering rate
%and $T_{c0}$ is the critical temperature of a clean sample),
%and the penetration depth at $T=0$, $\lambda (0)$.

For a conventional $s$-wave SC the effect of  non-magnetic
impurities on the temperature dependence of $\lambda (T)$ is small
and mostly irrelevant for all $T$. For $s^+$ superconductors, the
situation is qualitatively different because inter-band impurity scattering
$\Gamma_{\vpi}$
mixes hole and electron states with opposite values of the
order parameter $\pm\Delta$ and in this respect should be pairbreaking
and act in the same way as a magnetic impurity in a
conventional $s$-wave superconductor. Consequently, scattering
by non-magnetic impurities in $s^+$ SC affects $T_c$, the density
of states, {\it and} the temperature dependence of the penetration
depth. Over some range of $\Gamma_{\vpi}/\Delta$,
the behavior at the lowest $T$ is still exponential, however when
$\Gamma_{\vpi}/\Delta$ becomes larger than a critical value, superconductivity becomes gapless, and the exponential behavior disappears.

{\it Method.}~~~ The London penetration depth $\lambda (T)$ scales as
$1/\sqrt{\rho_s (T)}$, where $\rho_s (T)$ is the superfluid
density. The latter is, up to a factor, the zero frequency value
of the current-current correlation function and can be written in
the form\cite{rhosdef}
\be
\frac{\rho_s(T)}{\rho_{s0}} = \pi T \sum_{m} \frac{\tdm^2}{(\tdm^2 + \tem^2)^{3/2}} \,,
\label{eq:rho}
\ee
where $\rho_{s0}$ is the superfluid density at $T=0$ in the absence of
impurities. The integrand in \req{eq:rho} is defined in terms of
impurity-renormalized Matsubara energy, $\tem$, and the
superconducting vertex $\tdm$.
In an $s^+$ superconductor
the order parameters on the hole ($c$) and electron ($f$)
FS pockets are related,
$\tilde \Delta^c_m=-\tilde \Delta^f_m= \tilde \Delta_m$
and in Born approximation
\begin{subequations}\label{eq:temtdm}
\bea
i\tem &= & i\vare_m-\Gamma_{\bm 0}g^c(\tem,\tdm)-\Gamma_{\vpi}g^f(\tem,\tdm) \,,
\label{eq:tem}\\
\tdm &= &\Delta+\Gamma_{\bm 0}f^c(\tem,\tdm)+\Gamma_{\vpi}f^f(\tem,\tdm) \,,
\label{eq:tdm}
\eea
\end{subequations}
where $\vare_m = \pi T (2m +1)$, $\Gamma_{\bm 0} = \pi n_i N_F |u_0|^2$
and $\Gamma_{\vpi} = \pi n_i N_F |u_\pi|^2$ are the intra- and inter-band
impurity scattering rates, respectively
($u_{0,\pi}$ are impurity scattering amplitudes with
correspondingly small, or close to $\vpi=(\pi,\pi)$, momentum transfer),
$\Delta$ is the SC order parameter, and
functions $g^{c,f}$ and $f^{c,f}$ are
$\xi-$integrated normal and
anomalous Green's functions for holes and electrons:
\be
g^{c} = g^f = \frac{-i\tem}{\sqrt{\tem^2+\tdm^2}},\quad
f^{c} = - f^f = \frac{\tdm}{\sqrt{\tem^2+\tdm^2}}
\label{eq:gfdef}
\ee
Since the $f$-function has opposite signs in two bands,
 $\Gamma_{\vpi}$ has the same effect on anomalous
self-energy as the scattering on
 magnetic impurities in an ordinary $s$-wave superconductor.
Following the customary path one may introduce
$\eta_m=\tem/\vare_m$ and $\barD = \tdm/\eta_m$ that satisfy
\begin{subequations}
\bea
\eta_{m} = 1 + \left(\Gamma_{\bm 0} +
\Gamma_{\vpi}\right)\frac{1}{\sqrt{\barD^2 + \vare_m^2}};
\label{eq:etam}
\\
\barD = \Delta(T)  - 2 \Gamma_{\vpi}
\frac{\barD}{\sqrt{\barD^2 + \vare_m^2}}.
\label{eq:barD}
\eea
\end{subequations}
The order parameter $\Delta (T)$ is
determined by the self-consistency equation
\begin{equation}
\Delta(T) = V^{sc} \, \pi T \sum_{\vare_m}^{\Lambda} f^c(\tem,\tdm)=
\pi T \sum_{\vare_m}^\Lambda
\frac{V^{sc} \barD } {\sqrt{\barD^2 + \vare_m^2}},
\label{eq:selfcons}
\end{equation}
where  $V^{sc}$ is  the $s^+$  coupling constant and $\Lambda$ is the
ultraviolet cutoff.  Notice that the last expression contains $\barD$ and
\emph{bare} Matsubara frequencies $\vare_m$. Eqs. (\ref{eq:temtdm})-(\ref{eq:selfcons})
can be extended to the case when the gaps on hole and electron FS
have different magnitudes.

Solutions of the system of Eqs.~(\ref{eq:barD}) and (\ref{eq:selfcons}) give
the values of $\Delta(T)$ and $\barD$. In particular, \req{eq:barD} is an
algebraic equation (valid at any $T$) which expresses $\barD$ in terms of
$\Delta$. The latter itself depends on $\Gamma_{\vpi}$,  because the
self-consistency equation (\ref{eq:selfcons}) contains $\barD$.
 Without inter-band scattering ($\Gamma_{\vpi} =0$) we have
$\barD=\Delta= \Delta_0 =1.76 \; T_{c0}$, where $T_{c0}$ and $\Delta_0$ are the
BCS transition temperature and the $T=0$ gap in a clean superconductor.  For
$\Gamma_{\vpi} \neq 0$, ${\bar \Delta}_m$ differs from $\Delta$, and $\Delta$
differs from $\Delta_0$.  Converted to real frequencies, Eqs.~(\ref{eq:barD})
and (\ref{eq:selfcons}) yield  a complex function ${\bar \Delta} (\vare)$.
For $2 \Gamma_{\vpi} \geq \Delta$, ${\bar \Delta} (\vare =0)$ vanishes, i.e.,
superconductivity becomes gapless.\cite{abrikosov} At the critical point
$2 \Gamma_{\vpi} =\Delta$, ${\bar \Delta} (\vare) \propto (-i\vare)^{2/3}$ at small
$\vare$, at larger $\Gamma_{\vpi}$,  ${\bar \Delta} (\vare) = -i \; const \; \vare +
O(\vare^2)$.

\begin{figure}[t]
\centerline{\includegraphics[width=\linewidth]{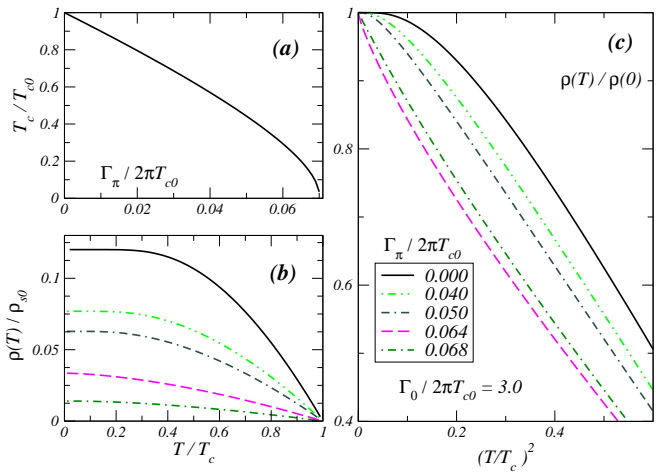}}
\caption{(color online) (a) Suppression of $T_c$ by inter-band scattering in a
two-band ($\Delta, -\Delta$) model; (b) superfluid stiffness $\rho_s (T)$ in a dirty $s^+$
superconductor for fixed intra-band impurity scattering $\Gamma_0/2\pi T_{c0} =3$,
and various inter-band scatterings $\ag = \Gamma_{\vpi}/2\pi T_{c0}$;
%(b) Log-Log plot showing power-law behavior for $\ag \gtrsim 0.06$;
and (c) low-$T$ plot of $\rho_s$ vs. $T^2$ showing near $n=2$ power-law around
onset of gapless regime.
}
\label{fig:1}
\end{figure}

{\it Results.}~~~
We  express the results using dimensionless parameter $\ag = \Gamma_{\vpi}/2\pi T_{c0}$. For equal gap magnitudes and $2 \Gamma_{\vpi}/\Delta <1$,
 $y = \Delta/\Delta_0$ is
the solution of $ y = \exp[-\pi e^\gamma \ag/y]$, where $\gamma
\approx 0.577$ is the Euler constant.\cite{skalski}  At a given $T$
a gapless superconductivity emerges, when $y$ becomes smaller
than $4 \ag e^\gamma$, i.e., for $\ag > (1/4) \exp[-(\gamma + \pi/4)]
\approx 0.064$. The transition temperature obeys~\cite{abrikosov,skalski}
$\ln(T_c/T_{c0}) = \Psi\left(1/2\right) - \Psi\left(1/2 + 2 \ag T_{c0}/T_{c}\right)$,
where $\Psi(x)$ is the di-Gamma function, Fig.~\ref{fig:1}a. $T_c$ decreases with $\ag$ and
vanishes at $\ag_{cr}  = e^{-\gamma}/8 \approx 0.07$
($\Gamma_{\vpi}/\Delta_0 = 1/4$). For $0.064 < \ag < \ag_{cr}$,
${\bar \Delta} (T,\vare) \propto i\vare$ for small $\vare$, including $T=0$,
and thus even $T =0$ zero-energy density of states becomes finite.
(At the onset, at $\ag =0.064$, $T_c \approx 0.22 \, T_{c0}$,
and $\Delta(0) = 0.46 \, \Delta_0$).
The ratio $2\Delta (0)/T_c$ increases with $\ag$ and reaches $7.2$ at the
onset of the gapless behavior and $8.88$ at $\ag = \ag_{cr}$
(Ref. \onlinecite{skalski}). A large value of $2\Delta(0)/T_c$ is often attributed to
strong coupling,\cite{marsiglio} but, as we see, can also be due to impurities.

%%%%%%%%%%%%%%%%%%%%%%%%%%%%%%%%%%%%%%%%%%%%%%%%%%%%%%%%%%%%%%%%%%%%%
%%%%%%%%%%%%%%%%%%%%%%%%%%%%%%%%%%%%%%%%%%%%%%%%%%%%%%%%%%%%%%%%%%%%%

In terms of auxiliary $\barD$ and $\eta_m$,
\be
\frac{\rho_s(T)}{\rho_{s0}} = \pi T \sum_{\vare_m}
\frac{{\barD}^2}{\eta_m(\barD^2 + \vare^2_m)^{3/2}}.
\label{3}
\ee
In general, the value of $\rho_s (T=0)$ and the functional form of
$\rho_s (T)$ depend on both $\Gamma_{\vpi}$ and $\Gamma_0$ because
$\Gamma_0$ is explicitly present in Eq.~(\ref{3})
via $\eta_m$ given by Eq.~(\ref{eq:etam}).
Impurity scattering amplitude is a decreasing function of momentum transfer,
and, in general,
$\Gamma_0 \gg \Gamma_{\vpi}$. Since we are interested in $\Gamma_{\vpi}
\sim \Delta$, we have $\Gamma_0 \gg \Delta$ and
\be
\rho_s (T) \approx B T \sum_{\vare_m} \frac{{\barD}^2}{\barD^2 + \vare^2_m},
\label{3_11}
\ee
where $B = \pi \rho_{s0}/(\Gamma_0 + \Gamma_{\vpi})$.
We see that $\Gamma_0$ only affects the overall factor $B$, and all
non-trivial $T$ dependence comes from frequency and temperature
dependence of ${\bar \Delta}_m$.

Several results for $\rho_s (T)$ given by (\ref{3_11})
can be obtained analytically.
First, near $T_c$,  $\rho_s (T) \propto \Delta^2 (T) \propto T_c -T$, i.e.,
\be
\frac{\rho_s}{\rho_s (T=0)} = B(\ag) \left (1-\frac{T}{T_c}\right),
\ee
where $\rho_s (T=0)$ is the actual zero-temperature value of $\rho_s$.
In a clean BCS superconductor $B=2$.
In the present dirty case ($\Gamma_0 \gg T_c, \Gamma_\pi$)
$B(\ag)$ is non-monotonic in $\ag$ and equals
$B(\ag \rightarrow  0) \approx 2.65,~B(\ag =0.064) \approx 1.67,~B(\ag \approx \ag_{cr}) = 2.03$.
This implies that a linear extrapolation of $\rho_s$
from $T \approx T_c$ to $T=0$ still yields a significantly
larger value than the actual $\rho_s (0)$.
Second, at $\ag < 0.064$, the $T$ dependence of
$\rho_s (T)$ remains exponential at low $T$,
$\rho_s (T) \propto e^{-{\bar \Delta} (\vare =0)/T}$ with
$\bar \Delta(\vare =0)=\Delta_0[1-(\ag/\ag_{cr})^{2/3}]^{3/2}$,
but at the onset of gapless superconductivity, when
${\bar\Delta} (\vare) \propto (-i \vare)^{2/3}$,
we have $\rho_s (T) \propto T^{5/3}$.
Finally, in the gapless regime $0.064 < \ag< \ag_{cr}$,
we found $\rho_s (T) \propto T^2$ at low $T$.

To obtain $\rho_s (T)$ at arbitrary $T$,
we self-consistently solved the gap equation
(\ref{eq:selfcons}) together with impurities (\ref{eq:temtdm})
and Green's functions (\ref{eq:gfdef})
numerically, found $\Delta(T)$ and $\tilde\Delta_m$, substituted them
into Eq. (\ref{eq:rho}) and obtained $\rho_s (T)$.  We present the results
in Fig.~\ref{fig:1} for several values of $\ag$.

We see that, once the inter-band impurity scattering increases, the range of
exponential behavior of $\rho_s (T)$ progressively shrinks to smaller $T$,
Fig.~\ref{fig:1}b.  Outside this low $T$-range, the  temperature dependence of
$\rho_s$ strongly resembles $T^2$ behavior, see Fig.~\ref{fig:1}(c).  The
$T^{5/3}$ behavior at the onset of gapless superconductivity is hard to see
numerically, as this power is confined to very low $T$, while for slightly
larger $T$ the behavior is again close to $T^2$.  Overall, the behavior of the
superfluid density in a relatively wide range of $\ag$ is a power-law $T^n$
with $n$ reasonably close to $2$ down to quite low $T$. At the same time, we
didn't find conditions under which the superfluid density would be linear at
low $T$.

%%%%%%%%%%%%%%%%%%%%%%%%%%%%%%%%%%%%%%%%%%%%%%%%%%%%%%%%%%%%%%%%%%%%%
\begin{figure}[t]
\centerline{\includegraphics[width=1.1\linewidth]{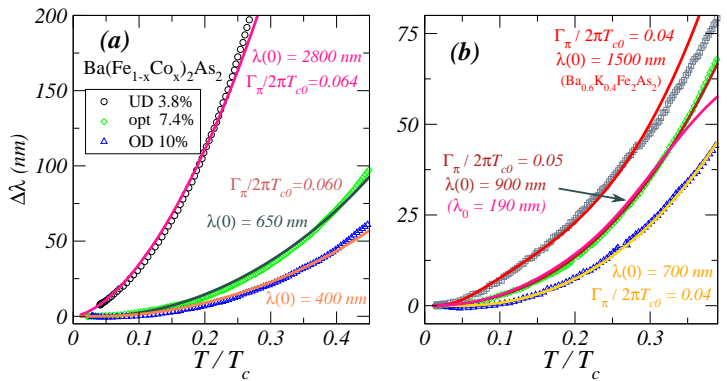}}
\caption{(color online) The fits to experimental data for
BaFe$_2$As$_2$~(Ref.\onlinecite{prozorov08}). % and LaFePO~(Ref.\onlinecite{bristol_1}).
We use only low-$T$ data as
at higher $T$  the experimental $\lambda (T)$ may be influenced
by sample geometry and fluctuations.
(a) The data for electron-doped BaFe$_2$As$_2$ for optimally doped
($x=7.4\%$) and overdoped ($x=10\%$) samples can be fitted reasonably well
using ($\Delta,-\Delta$)-model; % realistic $\lambda (0) \sim 400-700\, nm$
%(in the underdoped material $\lambda (T)$  is likely influenced by SDW magnetism).
(b) The fit of data for two Co-doped ($x=7.3\%,10\%$) and one K-doped samples
with a phenomenological extension of the presented two-band model to the case of
four gaps. We set gaps to be ($\Delta, -\Delta; \Delta, -\Delta_h$)
with $\Delta_h = \Delta/3\,,\, \Delta/2$.
In this case the pairbreaking parameter does not need to
be large.
The fitting values of $T=0$ penetration length are large but for 
$x=7.4\%$ sample we show a fit with inclusion of Fermi-liquid effects that 
reduces this parameter to $\lambda_{\mbox{\tiny FL}}(0) \equiv \lambda_0\sim 190$ nm, 
in agreement with experimental values. 
}
\label{fig:2}
\end{figure}
%%%%%%%%%%%%%%%%%%%%%%%%%%%%%%%%%%%%%%%%%%%%%%%%%%%%%%%%%%%%%%%%%%%%%

{\it Comparison with the data.}
Judging by the value of $2\Delta(0)/T_c$,~\cite{bristol} the material
with the least amount of inter-band impurity scattering is
SmFeAsO$_{1-x}$F$_x$, where $T_c \sim 55K$. In this compound it is difficult to expect
a large $\ag$, since the observed exponential BCS-like behavior of $\rho_s (T)$ at
small $T$ (Ref. \onlinecite{bristol}) is consistent with extended $s$-wave gap
and weak inter-band impurity scattering.\cite{YNagai08}

The data for electron- and hole-doped BaFe$_2$As$_2$ (Ref.~\onlinecite{prozorov08})
is fitted in Fig.~\ref{fig:2}a,b. The measured $\rho_s (T)$ scales approximately as $T^2$,
which is similar to behavior shown in Fig. \ref{fig:1}(c).
Left panel is the fit assuming that the gaps on two electron FSs and two hole FSs are
$\Delta,-\Delta$;
right panel is a more realistic fit in which we assumed, guided by ARPES
data,~\cite{diffgap} that the gaps on the inner hole and the two electron FS are
the same, but the gap on the other FS is 2-3 times smaller.

The values $\ag = 0.04 - 0.06$ used in these fits correspond to $T_c/T_{c0}
\sim 0.6-0.3$ which is consistent with  the values of $T_c \sim 10-30K$ in
this material, if we assume that $T_{c0}$ in the clean case is roughly the
same as in SmFeAsO. The curves shown in Figs. \ref{fig:2}a,b represent the
best fits, but we emphasize that we do not need to adjust $\ag$ to get a
$T^2$ behavior -- it persists over a range of $\ag$, see Fig.\ref{fig:1}c.
However we note that $\lambda(0)$ used in the fits is rather large
compared with the experimentally obtained values $\sim 200-300$ nm.\cite{prozorov08}
We suggest that this discrepancy may be due to the omission of
Fermi-liquid effects. 
The qualitative argument, supported by numerical estimates, is as follows.
Assume, by analogy with the cuprates, that fermion-fermion interactions
renormalize $\omega \to \omega Z_\omega$, where $Z_\omega$ is a decaying
function of frequency, and further assume that $Z_\omega \approx 1$ at energies
comparable to $\Delta$ 
so that it does not affect the relation between $\Delta(0)$ and $T_c$.
Then we find that the low-temperature dependence $f(T/T_c)$ of the
penetration length is rescaled
$\Del\lambda/\lambda_0 \sim f( Z_0 T/T_c)$ and the value of
the fitting parameter $\lambda_0\equiv\lambda_{\mbox{\tiny FL}}(0)$ 
decreases compared to what is obtained without Fermi liquid effects.
%our $\lambda (0) = \lambda_{meas} (0) * Z^2$, i.e., our
%$\lambda_0$ is larger than measured experimentally.
This $Z$ factor is
particularly relevant to heavily underdoped regime where it increases because
of SDW fluctuations. We believe this is the reason why we have to  use very
large $\lambda (0) =2800$nm to fit the data. At larger dopings, $Z$ is smaller,
but according to ARPES,\cite{ding} $Z \sim 2$ in
optimally doped Ba$_{1-x}$K$_x$Fe$_2$As$_2$. 
For $f \sim T^2$, from the analytic reasoning we get the effective $\lambda_0$
four times smaller than $\lambda(0)$,
reducing it to $\lambda_0 \sim 150 - 400$ nm,
in the range of what is experimentally extracted.
The numerical analysis confirms this and for $Z=2$ we show 
a fit for one of the electron-doped samples, that gives a reasonable 
value for the zero-temperature penetration length. 
%the $x=7.4\%$ sample. 
We therefore conclude that the penetration depth data for 122 material can be
fitted by a model of a dirty $s^+$ superconductor.  Note in this regard that
the data~\cite{bristol_3} show that $T_c$ is almost  insensitive to the value
of residual resistivity. This was interpreted as the argument for a
conventional $s$-wave gap. We note that this is also the case for extended
$s$-wave gap as the  dominant impurity scattering is intra-band scattering,
controlled by $\Gamma_0$, which affects residual resistivity but does not
affect $T_c$.

We attempted to fit the data for LaFePO using this approach but the fit fails
for all $\ag$ and one has to assume unrealistically large $\lambda (0)$ to
get even mediocre agreement with the data. From this perspective, it is
likely that the linear in $T$ behavior of $\lambda (T)$ in   LaFePO is not
related to impurities but rather is the consequence of the fact that the gap
in this material has nodes.

\emph{Conclusions.} In this paper we considered superfluid density $\rho_s (T)$ in a
multi-band superconductor with sign-changing $s^+$- symmetry in the presence of
non-magnetic impurities and applied the results to Fe-pnictides. We showed that
the behavior of the superfluid density is essentially the same as in an
ordinary $s$-wave superconductor with magnetic impurities. For a
moderate \emph{inter-band} impurity scattering, $\rho_s (T)$ over a wide range
of $T$ behaves roughly as $T^2$ and crosses over to exponential behavior only
at very low $T$. When superconductivity becomes gapless at $T=0$, the $T^2$
behavior extends to the lowest $T$. We argue that this power-law behavior is
consistent with recent experiments on penetration depth $\lambda (T)$ in hole
and electron doped BaFe$_2$As$_2$,
but we find that the present model
%%% is dirty $s^+$ superconductor model
does not explain the data for LaFePO.

Several modifications of the model may improve the comparison with
the experimental data.
First, an extension beyond Born approximation may further
flatten the $T$ dependence of $\lambda (T)$,
 although recent study for a different model\cite{bang09}
didn't find much changes beyond Born approximation.
Second, we assumed that the surface of a superconductor sample is homogeneous.
Defects on the surface may also modify the temperature
dependence of the penetration depth.

We acknowledge with thanks useful discussions with A. Carrington, I.~Eremin, R.
Prozorov and J. Schmalian. We thank R. Prozorov and A. Carrington for sharing
with us the data of Refs.~\onlinecite{prozorov08,bristol_1}
prior to publication.
This work  was supported by NSF-DMR 0604406 (A.V.Ch).

%%%%%%%%%%%%%%%%%%%%%%%%%%%%%%%%%%%%%%%%%%%%%%%%%%%%%%%%%%%%%%%%%%%%%%%%

\end{document}